\documentclass[journal,10pt]{IEEEtran}

\usepackage{cite}
\usepackage{graphicx}
\usepackage{amsmath}
\usepackage{array}
\usepackage{mdwmath}
\usepackage{amssymb}
\usepackage{mdwtab}
\usepackage{stfloats}
\usepackage{amsmath,amsthm}
\usepackage{threeparttable}
\usepackage{color}
\usepackage{url}
\usepackage{algpseudocode}
\usepackage{algorithm}
\usepackage{multicol}
\usepackage{amssymb}
\usepackage{epstopdf}
\usepackage[caption=false,font=footnotesize,labelfont=rm,textfont=rm]{subfig}
\algrenewcommand{\algorithmicrequire}{\textbf{Input:}}
\algrenewcommand{\algorithmicensure}{\textbf{Output:}}

\newtheorem{proposition}{Proposition}

\newtheorem{lemma}{Lemma}

\begin{document}

\title{\LARGE Elements Allocation for Joint Active and Passive IRS Aided Wireless Communications: A Rate-Maximization Perspective}

\author{Chaoying Huang, Wen Chen, Qingqing Wu, and Nan Cheng
 \vspace{-0.5cm}
\thanks{C. Huang, W. Chen, and Q. Wu are with the Department of Electronic Engineering, Shanghai Jiao Tong University, Shanghai 200240, China (e-mail: chaoyinghuang@sjtu.edu.cn; wenchen@sjtu.edu.cn; qingqingwu@sjtu.edu.cn).
}
\thanks{N. Cheng is with the School of Telecommunications Engineering, Xidian University, Xi’an 710071, China (e-mail: nancheng@xidian.edu.cn).
}
}

\markboth{}
{}

\maketitle
\begin{abstract}
Unlike previous works that focused solely on passive intelligent reflecting surface (PIRS) or active IRS (AIRS), a novel joint AIRS and PIRS architecture has been developed to flexibly utilize their combined advantages in mitigating multiplicative path loss cost-effectively. In this paper, we consider the AIRS-PIRS jointly aided wireless point-to-point communication system with two different deployment schemes in three-dimensional (3D) space. To balance the trade-off between the square-order beamforming gain of PIRS and the unique power amplification gain of AIRS, we optimize the elements allocation and beamforming design of the two IRSs under various practical constraints from a rate-maximization perspective. Moreover, we derive a series of element-related closed-form analytical expressions and compare the performance of the two schemes. Our analysis shows that in both schemes, PIRS should be allocated more elements than AIRS, and the received signal-to-noise ratio (SNR) increases asymptotically with the cube of the number of reflecting elements, when the distance between AIRS and PIRS is sufficiently large. Last, simulation results validate our analysis and indicate that both schemes can achieve superior rate performance over various benchmarks.

\end{abstract}
\begin{IEEEkeywords}
Intelligent reflecting surface (IRS), active IRS,
passive IRS, IRS elements allocation, rate maximization.
\end{IEEEkeywords}

\section{Introduction}

Intelligent reflecting surface (IRS) has emerged as a promising technology to reconfigure the wireless radio propagation environment via smart signal reflections \cite{wu2021intelligent}, \cite{wu2022irs}. Different from conventional active relays, a passive IRS (PIRS) is a low-cost metasurface that provides passive beamforming gain and operates in full-duplex mode \cite{zhang2020capacity}, \cite{kang2022irs}. Moreover, as it is beneficial to divide one single IRS into multiple smaller-size IRSs \cite{kang2022irs}, \cite{han2022double} investigated a double-PIRS system, where two PIRSs were allocated equal numbers of elements, achieving better performance than a single PIRS. In \cite{Mei2022csi}, two or more PIRSs were employed to assist each wireless link by jointly exploiting their single as well as multiple signal reflections. However, conventional PIRS still suffers from severe product-distance path loss resulting in radical limitation of power of multi-reflection signals \cite{long2021active}.

Recently, the active IRS (AIRS) has been proposed to efficiently compensate the PIRS issue by enabling simultaneous signal reflection and amplification which leads to an appreciable amplification power gain \cite{li2023active}. Note that since the AIRS is equipped with massive active elements manipulated by reflection-type amplifiers, it non-negligibly incurs amplification noise \cite{long2021active}. Generally, existing studies on IRS have revealed that AIRS and PIRS offer complementary advantages. To be specific, the AIRS-assisted wireless system has a lower beamforming gain of $\mathcal{O}(M)$ with $M$ reflecting active elements \cite{you2021wireless} than the PIRS case (i.e., $\mathcal{O}(M^2)$ \cite{wu2021intelligent}), which makes PIRS more appealing when $M$ is large. In contrast, AIRS outperforms PIRS when amplification power is high or $M$ is small, or when the power budget is not very small with the same power budget \cite{you2021wireless}. It follows that under a fixed deployment budget (both the total element budget and the amplification power budget), the conventional IRS with either passive or active elements only may not achieve the optimal system performance. 

Motivated by this idea, \cite{fu2023active} proposed a novel joint AIRS and PIRS architecture that combines the advantages of both IRSs and investigated their placement optimization in the two-dimensional (2D) plane. In \cite{fu2023multi}, the authors further studied the optimal location of the AIRS with the other PIRSs being fixed in the multi-AIRS/PIRS-enabled wireless networks. Moreover, a joint optimization problem of AIRS and PIRS positions and the number of their reflecting elements with the given maximum size of each IRS was formulated and solved in \cite{fu2024indoor} to minimize the deployment cost in a given indoor region. However, determining the size (i.e., number of elements) of each IRS to trade off between the square-order beamforming gain of PIRS and the unique power amplification gain of AIRS under a fixed deployment budget is still confusing. Specifically, when the size of AIRS increases and thus that of PIRS decreases, the passive beamforming gain consequently diminishes, and the amplification noise is non-negligibly expanded. Conversely, if the size of AIRS is too small, the amplification gain is insufficient. In addition, for scenarios where the positions of the transmitter (Tx) and receiver (Rx) are fixed, such as the fixed-location machines of the AIRS-PIRS assisted internet of things (IoT), deploying the optimal number of elements at each IRS is conducive to the communication rate improvement.

To this end, we focus on optimizing the elements allocation as well as the beamforming design in the AIRS-PIRS jointly aided wireless point-to-point communication system with a fixed deployment budget. In particular, we formulate the rate maximization problem by optimizing the elements allocation the beamforming design of the two IRSs for two transmission schemes, i.e., $\mathrm{Tx }\!\to\! \mathrm{AIRS}\!\to\!\mathrm{PIRS}\!\to\!\mathrm{Rx}$ (TAPR) and $\mathrm{Tx  }\!\to \!\mathrm{PIRS}\!\to\!\mathrm{AIRS}\!\to\!\mathrm{Rx}$ (TPAR) schemes. By introducing slack variables, we transform these non-convex problems into convex ones and derive the optimal solutions. Furthermore, we propose the alternating optimization (AO) algorithm to simultaneously obtain the optimal elements allocation and optimal placements of the two IRSs in three-dimensional (3D) space. To gain more useful insights, we provide the element-related closed-form expressions for the near-optimal solutions, which reveal that the PIRS should be allocated more elements than the AIRS and that the considered system attains cube SNR scaling order. Moreover, we take the achievable rates of the TAPR and TPAR schemes into comparison and simulation results show that both schemes can achieve superior rate performance over the benchmark systems.

\begin{figure}[t]
  \centering
  \includegraphics[width=8cm]{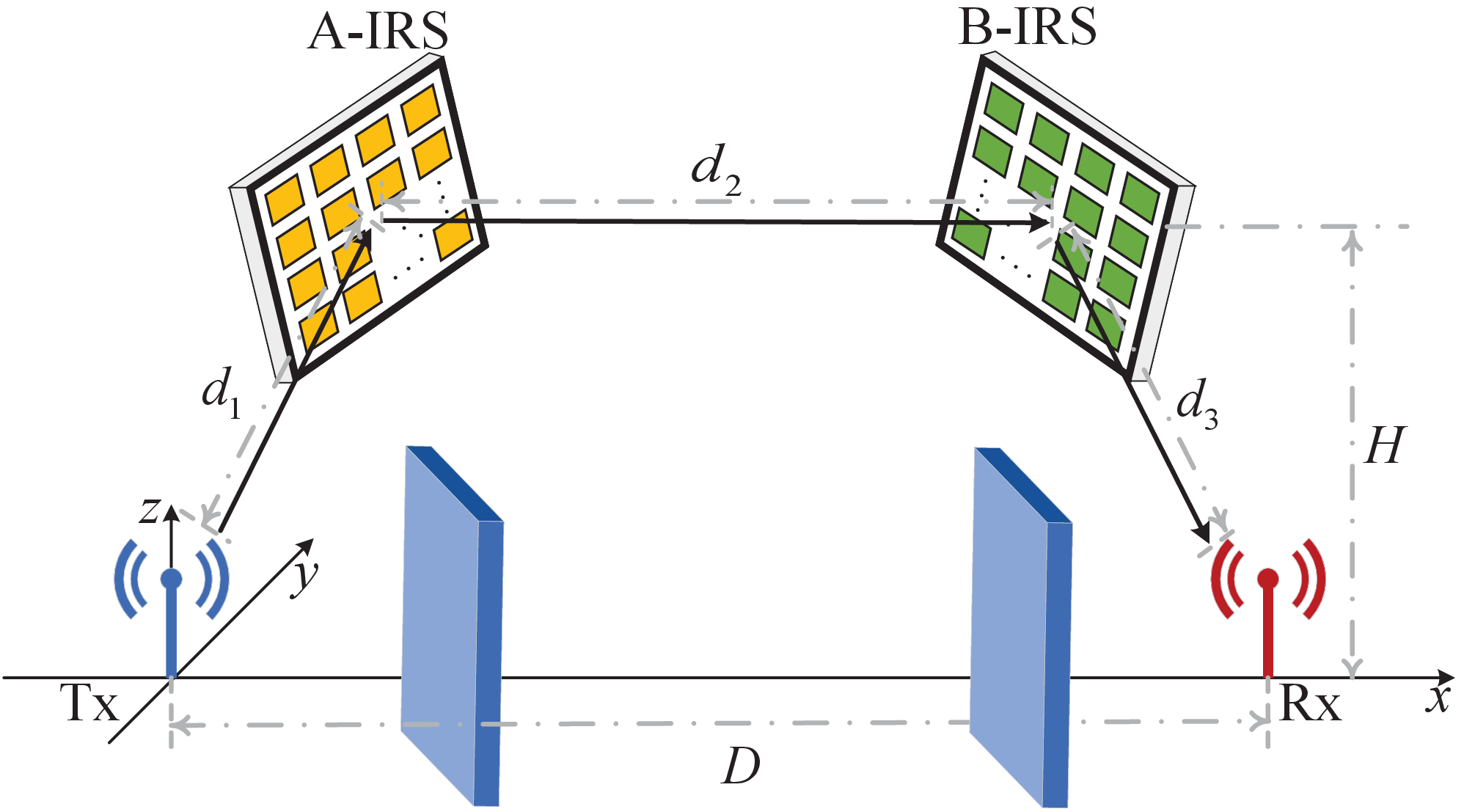}\\
  \caption{\small{The A-IRS and B-IRS jointly aided wireless communication system.}}\label{system}
  \vspace{-0.2cm}
\end{figure}

\section{System Model and Problem Formulation}
As illustrated in Fig.\ref{system}, we consider a double-IRS-aided wireless communication system in which active and passive IRSs equipped with appropriate elements jointly assist communication between a single-antenna\footnote{For the multiple-antenna case, the maximal ratio transmission/combining techniques can be extended to the Tx/Rx.} Tx and a single-antenna Rx\footnote{This work can be extended to multi-user scenarios, but further investigation on multi-access schemes is required. [8] have initially studied AIRS-aided multiple access designs.}. Without loss of generality, we assume that the first IRS and second IRS are expressed as A-IRS and B-IRS, respectively. Consider a 3D Cartesian coordinate system, where the Tx, Rx, and two IRSs are located at $u_{T}\!=\!(0,0,0)$, $u_{R}\!=\!(D,0,0)$, $u_{A}\!=\!(x_A,y_A,H)$ and $u_{B}\!=\!(x_B,y_B,H)$, respectively. Assuming that the direct Tx-Rx link is blocked by arbitrary obstacles, the data is transmitted only via a double-reflection link over three line-of-sight (LOS) channels.  

We denote that the two cascaded IRSs are equipped with $N_A$ and $N_B$ reflecting elements, respectively. Specifically, we define $\mathbf{\Psi } _{A}\!=\!\mathrm{diag} (\alpha _{1}e^{j\varphi _{1} },...,\alpha _{N_{A} }e^{j\varphi _{N_{A} } }  ) $ and $\mathbf{\Phi  } _{B}\!=\!\mathrm{diag} (\beta  _{1}e^{j\phi  _{1} },...,\beta _{N_{B} }e^{j\phi_{N_{B} } }  )$ as the reflection matrices of A-IRS and B-IRS respectively, where $\alpha _{n }$/$\beta _{n }$ and $\varphi _{n } $/$\phi _{n} $ correspondingly denote the reflection amplitude and phase shift. Then, we assume that the two IRSs are located in the far field of the Tx and Rx, so that the common reflection amplitude factors of A-IRS and B-IRS elements are set as $\alpha _{n}=\alpha\ge  1,\forall n\in \mathcal{N} _{A} \triangleq \left \{ 1,...,N_{A} \right \} $ and $\beta  _{n}=\beta \ge  1,\forall n\in \mathcal{N} _{B}\! \triangleq \left \{ 1,...,N_{B} \right \} $, respectively. Moreover, let $M$ represent the total element budget (i.e., total number of reflecting elements) with $W_A$ and $W_B$ respectively denoting the deployment cost of one A-IRS element and one B-IRS element. 

Denote the distance between Tx and A-IRS, A-IRS and B-IRS, and B-IRS and Rx as $d_1$, $d_2$ and $d_3$, respectively. Let $\boldsymbol{g}\!\in\! \mathbb{C} ^{N_{A}\times 1 } $, $\boldsymbol{S}\!\in\! \mathbb{C} ^{N_{B}\times N_{A}}  $, and $\boldsymbol{h}\! \in\!\mathbb{C} ^{N_{B}\times \!1} $ represent the baseband-equivalent LOS channel from the Tx to A-IRS, from the A-IRS to B-IRS, and from the B-IRS to Rx, respectively. Further, the Tx$\to $A-IRS link is expressed as $\boldsymbol{g}\!=\!\sqrt{\rho } /d_{1} e^{\frac{-j2\pi d_{1} }{\lambda } } \boldsymbol{{u}}_\mathrm{TA}(\theta_\mathrm{TA},\vartheta _\mathrm{TA},N_{A})$, where $\boldsymbol{{u}}_\mathrm{TA}(\theta_\mathrm{TA},\vartheta _\mathrm{TA},N_{A})\!=\!\boldsymbol{a}(\frac{2d_{1} }{\lambda } \sin (\theta _\mathrm{TA}) \sin(\vartheta _\mathrm{TA} ),N_{A,x}  )\otimes \boldsymbol{a}(\frac{2d_{1} }{\lambda } \cos (\vartheta _\mathrm{TA} ),N_{A,y} ),N_{A}\! =\!N_{A,x}\!\times\! N_{A,y}$ and $\boldsymbol{a}(w,N)\!=\!\left [ 1,...,e^{-j\pi (N-1)w}  \right ]^{T}   $. $\theta_{TA}$ and $\vartheta _{TA} $ are the azimuth and elevation angles of arrival at the A-IRS, respectively. $\lambda$ and $\rho\!=\!(\lambda /4\pi )^{2} $ denote the wavelength and the channel power gain at the reference distance of 1m, respectively. Similarly, the A-IRS$\to $B-IRS link is given by $\!\boldsymbol{S}\!=\!\sqrt{\rho } /d_{2} e^{\frac{-j2\pi d_{2} }{\lambda } } \boldsymbol{{u}}_\mathrm{AB}(\theta_\mathrm{AB},\vartheta _\mathrm{AB},N_{B})\boldsymbol{{u}}_\mathrm{BA}^{H}(\theta_\mathrm{BA},\vartheta _\mathrm{BA},N_{A})  $, and the B-IRS$\to $Rx link is given by $\boldsymbol{h}\!=\!\sqrt{\rho } /d_{3} e^{\frac{-j2\pi d_{3} }{\lambda } } \boldsymbol{{u}}_\mathrm{RB} (\theta_\mathrm{RB},\vartheta _\mathrm{RB},N_{B})$. In this letter, we assume the channel state information (CSI) is available.\footnote{The existing channel estimation techniques for double-IRS aided system proposed by \cite{Mei2022csi} were applicable to obtain the CSI in the current work.}

We aim to maximize the achievable rate $R$ among Rx of the above TAPR and TPAR transmission schemes by optimizing the element allocations of A-IRS and B-IRS, $\left \{ N_{A}, N_{B}\right \} $ and beamforming designs (i.e., reflection matrices) of A-IRS and B-IRS, $\left \{ \Psi _{A}, \Phi _{B}\right \} $. Given $N_{A}$ and $N_{B}$, the optimal phase shifts of double-IRS are expressed as \cite{fu2023active} 
\begin{align}\label{eq1}
\varphi _{n}= \mathrm{arg } \left ( \left [ \boldsymbol{u}_\mathrm{BA}   \right ]_{n}   \right ) -\mathrm{arg } \left ( \left [ \boldsymbol{u}_\mathrm{TA}   \right ]_{n}   \right ),\forall n\in \mathcal{N}_{A},
\end{align}
\begin{align}\label{eq2}
\phi  _{n}= \mathrm{arg } \left ( \left [ \boldsymbol{u}_\mathrm{AB}   \right ]_{n}   \right ) -\mathrm{arg } \left ( \left [ \boldsymbol{u}_\mathrm{RB}   \right ]_{n}   \right ),\forall n\in \mathcal{N}_{B}.
\end{align}

As such, our objective reduces to optimizing the number of A-IRS and B-IRS elements, $\left \{ N_{A}, N_{B}\right \} $, and the reflection amplitude factors of A-IRS and B-IRS, $\left \{ \alpha, \beta  \right \}$, for the achievable rate maximization in the TAPR and TPAR schemes under limited deployment budgets.

\subsection{TAPR Transmission Scheme}
We first consider the TAPR deployment scheme with $N_A\!=\!N_\mathrm{act}$, $N_B\!=\!N_\mathrm{pas}$, $W_A\!=\!W_\mathrm{act}$, $W_B\!=\!W_\mathrm{pas}$, $\boldsymbol{h}=\boldsymbol{h} _{\mathrm{RP} } $, $\boldsymbol{S}=\boldsymbol{S}_{\mathrm{AP} }$ and $\boldsymbol{g}=\boldsymbol{g}_{\mathrm{TA} }$, where $N_\mathrm{act}$ and $N_\mathrm{pas}$ denote the number of AIRS and PIRS elements respectively. Since the A-IRS is active and the B-IRS is passive in the TAPR system, we further set the reflection amplitude factor of the A-IRS as $\beta=1$ and that of the B-IRS is still constrained by $\alpha\ge 1$. 

\subsubsection{System Model} 
According to the deployment strategy of TAPR scheme with $\mathrm{Tx  }\!\to\! \mathrm{AIRS}\!\to\!\mathrm{PIRS}\!\to\!\mathrm{Rx}$ transmission link, the  signal received at the Rx is obtained as
\begin{equation}\label{eq3}
y_{\mathrm{AP} }=\boldsymbol{h} _{\mathrm{RP} }^{H} \mathbf{\Phi}_{\mathrm{pas} } \boldsymbol{S}_{\mathrm{AP} } \mathbf{\Psi }_{\mathrm{act} }\boldsymbol{g}_{\mathrm{TA} }s+ \boldsymbol{h} _{\mathrm{RP} }^{H} \mathbf{\Phi}_{\mathrm{pas} } \boldsymbol{S}_{\mathrm{AP} } \mathbf{\Psi }_{\mathrm{act} }\boldsymbol{v}+n_{0},  
\end{equation}
where $s$ denotes the transmitted signal for Rx with power $P_\mathrm{t} $. $\boldsymbol{v}\in \mathbb{C}^{N_\mathrm{act}\times 1 } $ is the amplification noise introduced by active elements of AIRS, and it satisfies the independent circularly symmetric complex Gaussian (CSCG) distribution $\boldsymbol{v}\sim \mathcal{CN}  (\boldsymbol{0}_{N_{\mathrm{act} } } ,\sigma_{v }^2 \mathbf{I} _{N_{\mathrm{act} } })$ with the amplification noise power $\sigma_{v}^2$. Besides, $n_{0} \sim \mathcal{CN}  (0,\sigma_{0}^2  )$ is the additive white Gaussian noise (AWGN) at the Rx. Note that the amplification noise passes through both the AIRS-PIRS and PIRS-Rx links to reach Rx in the TAPR scheme, suffering product-distance path loss. The corresponding received signal-to-noise ratio (SNR) of the receiver is given by 
\begin{equation}\label{eq4}
\gamma _{\mathrm{AP} } =\frac{P_\mathrm{t} \left | \boldsymbol{h} _{\mathrm{RP} }^{H} \mathbf{\Phi}_{\mathrm{pas} } \boldsymbol{S}_{\mathrm{AP} } \mathbf{\Psi }_{\mathrm{act} }\boldsymbol{g}_{\mathrm{TA} } \right |^{2}  }{\sigma _{v}^{2}\left \|  \boldsymbol{h} _{\mathrm{RP} }^{H} \mathbf{\Phi}_{\mathrm{pas} } \boldsymbol{S}_{\mathrm{AP} } \mathbf{\Psi }_{\mathrm{act} }\right \|^{2} +\sigma _{0}^{2}}.  
\end{equation}
Therefore, the achievable rate in bits per second per Hertz (bps/Hz) for the TAPR transmission system is
\begin{equation}\label{eq5}
R_{\mathrm{AP} } =\mathrm{log} _{2}(1+\gamma _{\mathrm{AP} }).
\end{equation}

\subsubsection{Problem Formulation} 
We aim to maximize the achievable rate for the Rx by jointly optimizing the elements allocation of the AIRS and PIRS $\left \{ N_\mathrm{act}, N_\mathrm{pas}\right \} $ and the reflection amplitude factor of the AIRS $\left \{ \alpha\right \} $ for the TAPR scheme, which is formulated as follows. 
\begin{align}
(\text{P1})\quad \underset{N_{\mathrm{act} },N_{\mathrm{pas} },\alpha }{\text{maximize}} \quad  & R_{\mathrm{AP} } \notag  \\
\mbox{s.t.}\quad
& W_{\mathrm{act} }N_{\mathrm{act} }+ W_{\mathrm{pas} }N_{\mathrm{pas} }\le M,\label{element budget constrain}\\
& P_{\mathrm{t} }\left \| \mathbf{\Psi}_{\mathrm{act} } \boldsymbol{g}_{\mathrm{TA} }   \right \|^{2}\!+\!\sigma _{v}^{2}\left \| \mathbf{\Psi}_{\mathrm{act} } \right \|^{2}\!\le \!P_{v} , \label{amplification factor constrain1}\\
& \alpha \ge 1,\label{amplification factor constrain2}\\
& N_{\mathrm{act} }  >  0, N_{\mathrm{pas} }  > 0, \label{element constrain}
\end{align}
where $P_{v}$ denotes the given amplification power budget of all AIRS elements. Constraint (6) indicates that the number of elements for the two IRSs is constrained by the total element budget. Meanwhile, constraint (7) limits the sum of transmitted signal amplification power and amplification noise power over the AIRS output to within the amplification power budget. Since the objective function increases with the amplification power factor $\alpha$, the equality in constraint (\ref{amplification factor constrain1}) always holds at the optimal solution of problem (P1). It is not difficult to verify that the optimal amplification factor with given $\left \{ N_{\mathrm{act} },N_{\mathrm{pas}} \right \} $ satisfies $\alpha ^{\ast }  =\sqrt{P_{v} d_{1}^{2}/(P_t\rho N_\mathrm{act} +d_{1}^{2}\sigma _{v}^{2}N_\mathrm{act})  }$.
Hence, the received SNR can be further expressed as $ \gamma _{\mathrm{AP} }=P_tP_v\rho ^{3}/\zeta_{\mathrm{AP} } $, where
\begin{equation}\label{eq6}
\zeta_{\mathrm{AP} }=\frac{P_v\sigma _{v}^{2}\rho ^{2}d_{1}^{2}}{N_{\mathrm{act} } } +\frac{\sigma _{0}^{2} d_{2}^{2}d_{3}^{2}(\rho P_t+\sigma _{v}^{2}d_{1}^{2})}{N_{\mathrm{act} }N_{\mathrm{pas} }^{2}}.
\end{equation}

Based on the above discussions, problem (P1) can be equivalent to the following problem (P2).
\begin{align}
(\text{P2})\quad \underset{N_{\mathrm{act} },N_{\mathrm{pas} }}{\text{minimize}} \quad  & \zeta _{\mathrm{AP} } \notag  \\
\mbox{s.t.}\quad
& \mathrm{Constraints} \enspace\mathrm{(6),(9)} . 
\end{align}
It is challenging to solve (P2) due to the integer constraint (\ref{element constrain}) and the non-convex objective function. To tackle this difficulty, we first relax the integer values $\left \{ {N_{\mathrm{act} },N_{\mathrm{pas} }} \right \}  $ into the continuous values $\left \{ {x_{\mathrm{act} },x_{\mathrm{pas} }} \right \}$. Then, we introduce the slack variables $\tilde{x} _{\mathrm{a} } = \mathrm{log}  (x_{\mathrm{act} })$ and $\tilde{x} _{\mathrm{p} } = \mathrm{log}  (x_{\mathrm{pas} })$, yielding the objective function transform to
\begin{equation}\label{eq7}
 \tilde{\zeta} _{\mathrm{AP} } =C_{1}^{\mathrm{AP}}e^{-\tilde{x}  _{a} }  +(C_{2}^{\mathrm{AP}}+C_{3}^{\mathrm{AP}}) e^{-\tilde{x}_{a}-2\tilde{x}_{p} },
\end{equation}
where $C_{1}^{\mathrm{AP}}=P_v\sigma _{v}^{2}\rho ^{2}d_{1}^{2} $, $C_{2}^{\mathrm{AP}}=\sigma _{0}^{2} d_{2}^{2}d_{3}^{2}\rho P_t$, and $C_{3}^{\mathrm{AP}}=\sigma _{0}^{2}\sigma_{v}^{2}d_{1}^{2}d_{2}^{2}d_{3}^{2}$.
Thus, (P2) is correspondingly expressed as 
\begin{align}
(\text{P2.1})\quad \underset{\tilde{x} _{\mathrm{a} },\tilde{x} _{\mathrm{p} }}{\text{minimize}} \quad  & \tilde{\zeta} _{\mathrm{AP} } \notag  \\
\mbox{s.t.}\quad
& W_{\mathrm{act} }e^{\tilde{x} _{a} }  + W_{\mathrm{pas} }e^{\tilde{x} _{p} }\le M.\label{M-constrain-t}
\end{align}
It can be derived that the constraint (\ref{M-constrain-t}) needs to be active for the optimal solution of (P2.1). Hence, we can demonstrate that (P2.1) is a convex optimization problem as it has a convex objective function and constraint, which can be optimally solved using the interior-point method. 
By rounding the optimal continuous solutions, the integer elements allocation of AIRS and PIRS can be reconfigured.



\subsection{TPAR Transmission Scheme}
Then, consider the TPAR scheme with $N_A\!=\!N_\mathrm{pas}$, $N_B\!=\!N_\mathrm{act}$, $W_A\!=\!W_\mathrm{pas}$, $W_B\!=\!W_\mathrm{act}$, $\boldsymbol{h}=\boldsymbol{h} _{\mathrm{RA} } $, $\boldsymbol{S}=\boldsymbol{S}_{\mathrm{PA} }$ and $\boldsymbol{g}=\boldsymbol{g}_{\mathrm{TP} }$. We set the amplification factors as $\alpha\!=\!1$ and $\beta\!\ge \!1$, since the A-IRS is passive and B-IRS is active in the TPAR transmission scheme. Furthermore, the amplification noise reaches Rx only through the AIRS-Rx link in the TPAR scheme. According to the placement order of the aforementioned two IRSs, the received signal at Rx for TPAR can be shown as
\begin{equation}\label{eq8}
y_{\mathrm{PA} }=\boldsymbol{h} _{\mathrm{RA} }^{H} \mathbf{\Phi}_{\mathrm{act} } \boldsymbol{S}_{\mathrm{PA} } \mathbf{\Psi }_{\mathrm{pas} }\boldsymbol{g}_{\mathrm{TP} }s+ \boldsymbol{h} _{\mathrm{RA} }^{H} \mathbf{\Phi}_{\mathrm{act} }\boldsymbol{v}+n_{0}. 
\end{equation}
Consequently, the corresponding SNR at Rx is obtained as
\begin{equation}\label{eq9}
\gamma _{\mathrm{PA} } =\frac{P_{t} \left | \boldsymbol{h} _{\mathrm{RA} }^{H} \mathbf{\Phi}_{\mathrm{act} } \boldsymbol{S}_{\mathrm{PA} } \mathbf{\Psi }_{\mathrm{pas} }\boldsymbol{g}_{\mathrm{TP} } \right |^{2}  }{\sigma _{v}^{2}\left \|  \boldsymbol{h} _{\mathrm{RA} }^{H} \mathbf{\Phi}_{\mathrm{act} }   \right \|^{2} +\sigma _{0}^{2}}.
\end{equation}
Similar to the amplification power constraint in TAPR, the optimal amplification factor satisfies $\beta  ^{\ast }\! \! =\!\!\sqrt{P_{v} d_{2}^{2}/(P_t\rho^{2}  x_\mathrm{act} x_\mathrm{pas}^{2}/d_{1}^{2}  \!+\!d_{2}^{2}\sigma _{v}^{2}x_\mathrm{act})}$. By substituting $\beta  ^{\ast }$ to (\ref{eq9}), we execute the SNR re-expression as $ \gamma _{\mathrm{PA} }=P_tP_v\rho ^{3}/\zeta_{\mathrm{PA} }$, where
\begin{equation}\label{eq10}
\zeta_{\mathrm{PA} }=\frac{P_t\sigma _{0}^{2}\rho^{2} d_{3}^{2}}{x_{\mathrm{act} } } +\frac{\sigma _{v}^{2} d_{1}^{2}d_{2}^{2}(\rho P_v+\sigma _{0}^{2}d_{3}^{2})}{x_{\mathrm{act} }x_{\mathrm{pas} }^{2}}.
\end{equation}
By introducing the slack variables $\tilde{x} _{\mathrm{a} } = \mathrm{log}  (x_{\mathrm{act} })$ and $\tilde{x} _{\mathrm{p} } = \mathrm{log}  (x_{\mathrm{pas} })$ similarly to TAPR, the equality (\ref{eq10}) can be re-shown as
\begin{equation}\label{eq11}
 \tilde{\zeta} _{\mathrm{PA} } =(C_{1}^{\mathrm{PA}}+C_{2}^{\mathrm{PA}})e^{-\tilde{x}_{a}-2\tilde{x}_{p} }  +C_{3}^{\mathrm{PA}} e^{-\tilde{x}  _{a} },
\end{equation}
where $C_{1}^{\mathrm{PA}}=\sigma _{v}^{2} d_{1}^{2}d_{2}^{2} \rho P_v$, $C_{2}^{\mathrm{PA}}=\sigma _{v}^{2} d_{1}^{2}d_{2}^{2}\sigma _{0}^{2}d_{3}^{2}$, and $C_{3}^{\mathrm{PA}}=P_t\sigma _{0}^{2}\rho^{2} d_{3}^{2} $. Thus, the optimization problem for TPAR is formulated as 
\begin{align}
(\text{P3})\quad \underset{\tilde{x} _{\mathrm{a} },\tilde{x} _{\mathrm{p} }}{\text{minimize}} \quad  & \tilde{\zeta} _{\mathrm{PA} } \notag  \\
\mbox{s.t.}\quad
& W_{\mathrm{act} }e^{\tilde{x} _{a} }  + W_{\mathrm{pas} }e^{\tilde{x} _{p} }\le M . \label{M constrain-TPAR}
\end{align}
It is not difficult to verify that the problem (P3) is convex, so that its optimal solution can be obtained by utilizing the interior-point method.

Note that although this paper focuses on the elements allocation optimization, the optimal placements of two IRSs can be simultaneously obtained by the AO algorithm in both schemes. To be specific, the following two steps are alternately and iteratively carried out until the maximum achievable rate is reached: 1) Given the optimal elements allocation from the previous iteration, the optimal placements of the two IRSs for the current iteration are obtained by exhaustive searching for $\boldsymbol{D}=\left \{ d_1,d_2,d_3 \right \} $ under practical distance constraints. 2) Given the optimal placements from the previous iteration, the optimal elements allocation for the current iteration can be obtained by the interior-point method.



\section{Low-Complexity Allocation Design and Performance Analysis}
In this section, we propose suboptimal solutions of elements allocation in closed-form for the TAPR and TPAR schemes, thereby reducing computational complexity and shedding meaningful insights.

\subsection{Suboptimal Solutions}
For simplifying (\ref{eq6}) and (\ref{eq10}), we first provide a helpful Lemma, the proof of which is ignored for brevity.
\begin{lemma}
We have $(C_{2}^{\mathrm{AP}}+C_{3}^{\mathrm{AP}})/x_{\mathrm{act}}x_{\mathrm{pas}}^{2}  \! \gg\! C_{1}^{\mathrm{AP}}/x_{\mathrm{act}} $ and $C_{1}^{\mathrm{PA}}/x_{\mathrm{act}}x_{\mathrm{pas}}^{2} \!\gg \!C_{2}^{\mathrm{PA}}/x_{\mathrm{act}}x_{\mathrm{pas}}^{2}+C_{3}^{\mathrm{PA}}/x_{\mathrm{act}} $, if
\begin{equation}\label{eq12}
\mathrm{max } (\frac{\sqrt{P_{v}\rho }\sigma _{v}d_{1}x_{\mathrm{pas} }}{\sqrt{P_{t}} \sigma _{0}d_{3}} ,\frac{\sqrt{P_{t}}\sigma _{0} \rho d_{3}x_{\mathrm{pas} } }{\sqrt{\sigma _{v}^{2}d_{1}^{2}(\rho P_{v}-\sigma _{0}^{2}d_{3}^{2})} } ) \ll   d_{2}.
\end{equation}
\end{lemma}

\subsubsection{TAPR Scheme}
When condition (19) is satisfied, the first term of the denominator of the SNR for the TAPR scheme, which is the first term of the objective function of problem (P2), can be ignored. As such, the achievable rate at the Rx for the TAPR can be approximated as $\bar{R} _{\mathrm{AP} }=\mathrm{log} _{2}(1+P_tP_v\rho ^{3}/\bar{\zeta} _{\mathrm{AP} }({x} _{\mathrm{act } },{x} _{\mathrm{pas} }) )$, where $\bar{\zeta} _{\mathrm{AP} }({x} _{\mathrm{act } },{x} _{\mathrm{pas} }) =\frac{\sigma _{0}^{2} d_{2}^{2}d_{3}^{2}(\rho P_t+\sigma _{v}^{2}d_{1}^{2})}{x_{\mathrm{act} }x_{\mathrm{pas} }^{2}}$ is the denominator of the approximate SNR of TAPR scheme. Accordingly, the problem (P2) after integer relaxation is expressed approximately as
\begin{align}
(\text{P5})\quad \underset{{x} _{\mathrm{act } },{x} _{\mathrm{pas} }}{\text{min}} \quad  & \bar{\zeta}  _{\mathrm{AP} } \notag  \\
\mbox{s.t.}\quad
& W_{\mathrm{act} }{x} _{\mathrm{act} }   + W_{\mathrm{pas} }{x} _{\mathrm{pas} } \le M \label{sub_M}, \\ 
& x_{\mathrm{act} }>  0,  x_{\mathrm{pas} }>  0 .\label{sub_act} 
\end{align}
The optimal solution to (P5) is shown as follows.
\begin{lemma}
The optimal solution to (P5) is
\begin{equation}\label{eq17}
\bar{x}  _{\mathrm{act} } ^{\ast } =\frac{M}{3W_{\mathrm{act} } } , \quad \bar{x} _{\mathrm{pas} } ^{\ast } =\frac{2M}{3W_{\mathrm{pas} }}.
\end{equation}
\end{lemma}

{\emph {Proof:}}
Note that the constraint (\ref{sub_M}) always satisfies equality in the optimal solution since the objective function decreases with $M$. We take the first-order derivative of $\bar{\zeta} _{\mathrm{AP} }$ over $x _{\mathrm{pas} }$ with $x _{\mathrm{act} } =\frac{M-W_{\mathrm{pas} } x_{\mathrm{pas} } } {W_{\mathrm{act} } } $ as  
\begin{equation}\label{eq18}
\frac{\mathrm{d} \bar{\zeta} _{\mathrm{AP} }}{\mathrm{d}x_{\mathrm{pas} }  } =\frac{-C_{2}^{\mathrm{AP}}f_{1}(x_{\mathrm{pas} } )}{(Mx_{\mathrm{pas}}^{2} -W_{\mathrm{pas} }x_{\mathrm{pas}}^{3} )^{2} },
\end{equation} 
where $f_{1}(x_{\mathrm{pas} } ) =2Mx_{\mathrm{pas} }-3W_{\mathrm{pas} }x_{\mathrm{pas}}^{2}$. Note that $f_{1}(x_{\mathrm{pas} } )$ is a quadratic function. Meanwhile when $x_{\mathrm{pas} }\!>\!0$, we have $f_{1}(x_{\mathrm{pas} } )\!>\!0$, when $x_{\mathrm{pas} }\!>\! M$, we have $f_{1}(x_{\mathrm{pas} } )\!<\!0$, there exists a unique root of $f_{1}(x_{\mathrm{pas} } )$ within $\left ( 0,M \right ) $, i.e., $x_{\mathrm{pas} } ^{\mathrm{rt} } \!=\!\frac{2M}{3W_{\mathrm{act} }}$. Furthermore, when $0\!<\!x_{\mathrm{pas} }\!<\!x_{\mathrm{pas} } ^{\mathrm{rt} }$, $\frac{\mathrm{d} \bar{\zeta} _{\mathrm{AP} }}{\mathrm{d}x_{\mathrm{pas} }  }\!<\!0$ holds, i.e., $\bar{\zeta} _{\mathrm{AP} }$ monotonically decreases with $x_{\mathrm{pas} } $. When $x_{\mathrm{pas} } ^{\mathrm{rt} }\!<\!x_{\mathrm{pas} }\!<\!M$, $\frac{\mathrm{d} \bar{\zeta} _{\mathrm{AP} }}{\mathrm{d}x_{\mathrm{pas} }  }\!>\!0$ holds, i.e., $\bar{\zeta} _{\mathrm{AP} }$ monotonically increases with $x_{\mathrm{pas} }$. Accordingly, $f_{1}(x_{\mathrm{pas} } )$ is minimized at $\quad x _{\mathrm{pas} } ^{\ast }\! =\!x_{\mathrm{pas} } ^{\mathrm{rt} }$, which completes the proof. 
$\hfill\blacksquare$

\begin{proposition}\label{pro1}
By suboptimally equipping AIRS and PIRS for a TAPR system with the condition (\ref{eq12}), deploying two cooperative IRSs with optimal phase shift given in (\ref{eq1}) and (\ref{eq2}) results in an $M^{3}$-fold scaling order of the received SNR, i.e.,
\begin{equation}\label{eq13}
\bar{\gamma}_\mathrm{AP}=\frac{4M^{3}P_t P_v\rho ^{3}  }{27\sigma _{0}^{2} d_{2}^{2}d_{3}^{2}(\rho P_t+\sigma _{v}^{2}d_{1}^{2}) W_{\mathrm{act} }W_{\mathrm{pas} }^2  }.
\end{equation} 
\end{proposition}
Note that Proposition \ref{pro1} is not difficult to prove by substituting (\ref{eq17}) into the SNR for TAPR, which can be explained as follows. As the received signal experiences double-reflection which contributes to the multiplicative beamforming gain, its power subsequently increases in the order of $\mathcal{O}(M^{4} )$. While the noise power mainly scales with the order of $\mathcal{O}(M)$ when $d_2$ is considerably large, engendering the received SNR scaling order of $\mathcal{O}(M^{3} )$ for the TAPR system, which is higher than the scaling for the single-active-IRS case (i.e., $\mathcal{O}(M )$ \cite{you2021wireless}) and the single-passive-IRS case (i.e., $\mathcal{O}(M^2 )$ \cite{wu2021intelligent}).

\subsubsection{TPAR Scheme}
Based on Lemma 1, the denominator of the SNR for the TPAR scheme, i.e., the equation (16), can be approximated as $\bar{\zeta} _{\mathrm{PA} }({x} _{\mathrm{act } },{x} _{\mathrm{pas} }) =\frac{ d_{1}^{2}d_{2}^{2}\sigma _{v}^{2}\rho P_v}{x_{\mathrm{act} }x_{\mathrm{pas}} ^{2} }$, and thus the achievable rate for TPAR can be simplified to $\bar{R} _{\mathrm{PA} }=\mathrm{log} _{2}(1+P_tP_v\rho ^{3}/\bar{\zeta} _{\mathrm{PA} }({x} _{\mathrm{act } },{x} _{\mathrm{pas} }) )$. Therefore, the optimization problem for the TPAR scheme is approximated as  
\begin{align}
(\text{P6}) \ \underset{{x} _{\mathrm{act } },{x} _{\mathrm{pas} }}{\text{min}} \ & \bar{\zeta}  _{\mathrm{PA} } \notag \\
\mbox{s.t.}\
 &\mathrm{Constrains}\  (\ref{sub_M}),(\ref{sub_act}).
\end{align}
\begin{lemma}
The optimal solution to (P6) is given by
\begin{equation}\label{eq19}
\bar{x}  _{\mathrm{act} } ^{\ast } =\frac{M}{3W_{\mathrm{act} } } , \quad \bar{x} _{\mathrm{pas} } ^{\ast } =\frac{2M}{3W_{\mathrm{pas} }}.
\end{equation}
\end{lemma}
The proof is similar to that of Lemma 2 and is omitted for brevity. It is interesting to note that the element allocations to the AIRS and PIRS among the TAPR and TPAR schemes tend to be the same and independent of the placement of the two IRSs, while the PIRS is allocated more elements than the AIRS.

\begin{proposition}\label{pro2}
By suboptimally equipping PIRS and AIRS for a TPAR system with the condition (\ref{eq12}), deploying two cooperative IRSs with optimal phase shift given in (\ref{eq1}) and (\ref{eq2}) results in an $M^{3}$-fold scaling order of the received SNR, i.e.,
\begin{equation}\label{eq20}
\bar{\gamma} _\mathrm{PA} =\frac{4M^{3}P_t \rho ^{3}  }{27d_{1}^{2}d_{2}^{2}\sigma _{v}^{2}\rho W_{\mathrm{act} }W_{\mathrm{pas} }^2}.
\end{equation} 
\end{proposition}
Obviously, Proposition 2 can be proved straightly by substituting (\ref{eq19}) into the received SNR for TPAR. Similar to the TAPR system, the received SNR scaling order of $\mathcal{O}(M^{3})$ is owing to the received signal power scaling order of $\mathcal{O}(M^{4})$ and the noise power scaling order of $\mathcal{O}(M)$ when $d_2$ is considerably large.

\subsection{TAPR Versus TPAR}
Based on the suboptimal AIRS and PIRS elements allocation given in (\ref{eq17}) and (\ref{eq19}), we take the achievable rate of the TAPR and TPAR systems into comparison as follows.
\begin{proposition}\label{pro3}
The achievable rate of TAPR is no less than that of TPAR, i.e., $\bar{R} _{\mathrm{AP} } \ge \bar{R} _{\mathrm{PA} }$, when
\begin{equation}\label{eq21}
\frac{1}{\rho } \le  \frac{P_v}{d_3^2\sigma _{0}^{2}} - \frac{P_t}{d_1^2\sigma _{v}^{2} },
\end{equation} 
with the condition in (\ref{eq12}). Otherwise, $\bar{R} _{\mathrm{AP} } < \bar{R} _{\mathrm{PA} }$.
\end{proposition}
{\emph {Proof:}}
As the achievable rate monotonically increases with the SNR, it is feasible to compare $\bar{\gamma}_\mathrm{AP} $ and $\bar{\gamma}  _{\mathrm{PA} } $ instead of $\bar{R} _{\mathrm{AP} }$ and $\bar{R} _{\mathrm{PA} }$, i.e.,
\begin{align}
\bar{\gamma}  _{\mathrm{AP} } -\bar{\gamma}  _{\mathrm{PA} } = & k(\frac{1}{\sigma _{0}^{2} d_{2}^{2}d_{3}^{2}(\rho P_t+\sigma _{v}^{2}d_{1}^{2})} -\frac{1}{d_{1}^{2}d_{2}^{2}\sigma _{v}^{2}\rho P_v }) \notag \\
= & \frac{k(\frac{P_v}{d_3^2\sigma _0^2} -\frac{P_t}{d_1^2\sigma _{v}^2 } -\frac{1}{\rho })}{P_vd_2^2(\rho P_t+\sigma _{v}^{2}d_{1}^{2})} ,
\end{align} 
where $k=\frac{4M^{3}P_t P_v\rho ^{3}  }{27 W_{\mathrm{act} }W_{\mathrm{pas} }^2}>0$. As such, we have $\bar{\gamma}  _{\mathrm{AP} } \ge\bar{\gamma}  _{\mathrm{PA} }$, i.e., $\bar{R} _{\mathrm{AP} } \ge \bar{R} _{\mathrm{PA} }$, when $\frac{1}{\rho } \le  \frac{P_v}{d_3^2\sigma _{0}^{2}} - \frac{P_t}{d_1^2\sigma _{v}^{2} }$ , which completes the proof. Note that the condition in (\ref{eq21}) can be practically satisfied when the AIRS amplification power is high and/or the distance $d_3$ is sufficiently small. 
$\hfill\blacksquare$

\begin{figure}[!t]
\centering
\subfloat[The number of A-IRS elements\\ versus total element budget.]{\label{M_opt_sub}
		\includegraphics[scale=0.3]{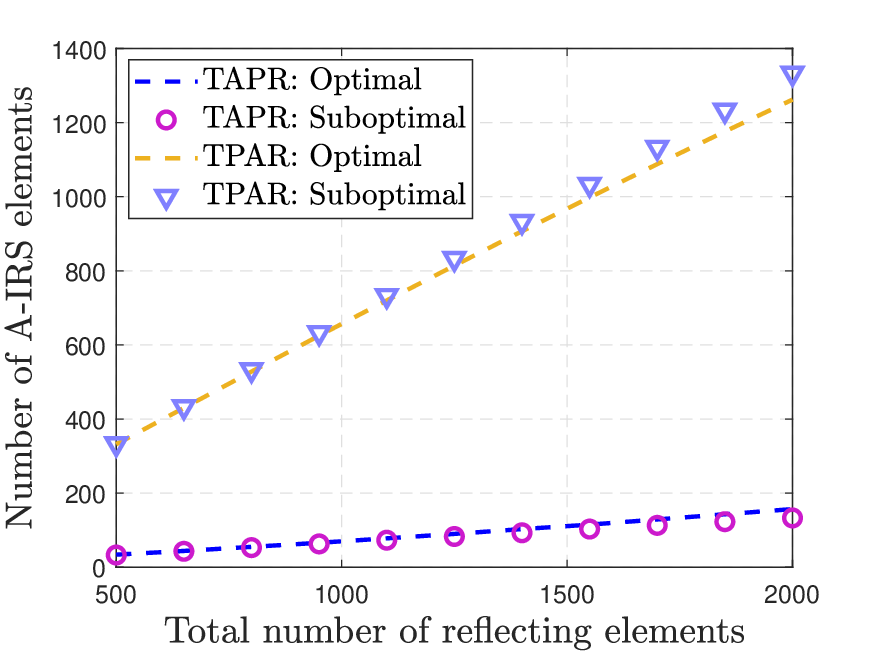}}
\subfloat[Achievable rate versus amplification power budget.]{\label{P_opt_sub}
		\includegraphics[scale=0.3]{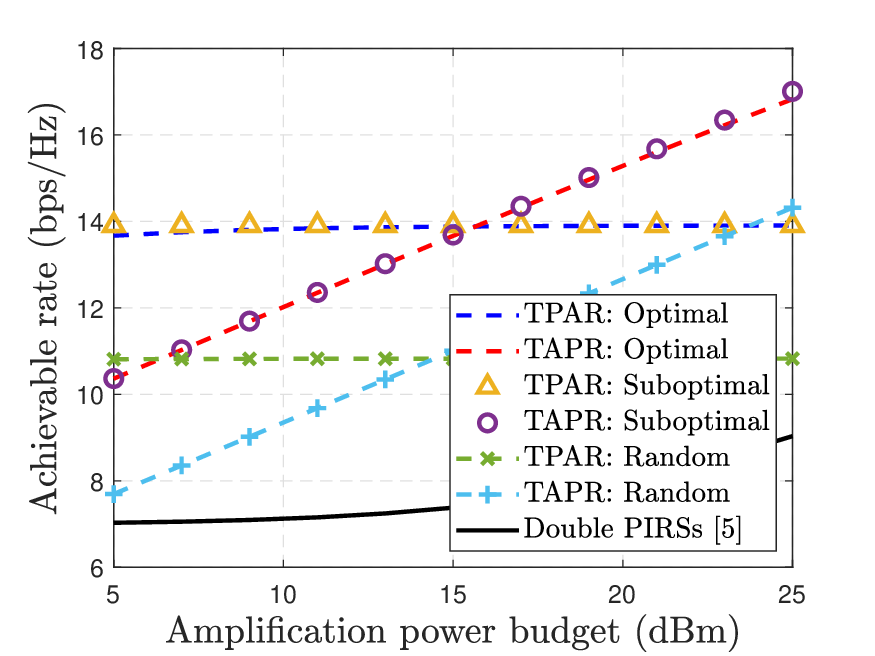}}
\caption{Effect of the total budget on elements allocation and achievable rate.}
\label{fig2}
\end{figure}

\section{Simulation Results}
In this section, numerical results are provided to compare the rate performance of the considered TAPR and TPAR transmission schemes with other benchmarks. The deployment locations of the Tx, the A-IRS, the B-IRS, and the Rx are set as (0,0,0), (15,5,10), (98,5,10) and (100,0,0) in meters (m), respectively. We set the channel power gain at the reference distance of 1m to $\rho=-30\mathrm{dB}$. Other parameters are set as $P_t\!=\!20\mathrm{dBm}$, $P_v\!=\!17\mathrm{dBm}$, $\sigma _{0}^2 \!=\!\sigma _{v}^2\!=\!-80\mathrm{dBm} $, $W_\mathrm{act}\!=\!5$ and $W_\mathrm{pas}\!=\!1$. $W_\mathrm{act}$ represents the deployment cost of each active element, which is larger than that of each passive element $W_\mathrm{pas}$ as AIRS bears more sophisticated hardware and higher static operation power. We consider four other systems as benchmarks, including a single PIRS (see \cite{wu2021intelligent}), a single AIRS (see \cite{long2021active}), a hybrid IRS (i.e., IRS consisted of passive and active reflecting elements \cite{kang2024hybrid}), as well as double PIRSs (see \cite{han2022double}). For the sake of fairness, we set the transmit power of the single PIRS and double PIRSs cases as ($P_t+P_v$). 

\begin{figure}[!t]
\centering
\subfloat[Achievable rate versus total ele-\\ment budget.]{\label{M_vs_R}
		\includegraphics[scale=0.3]{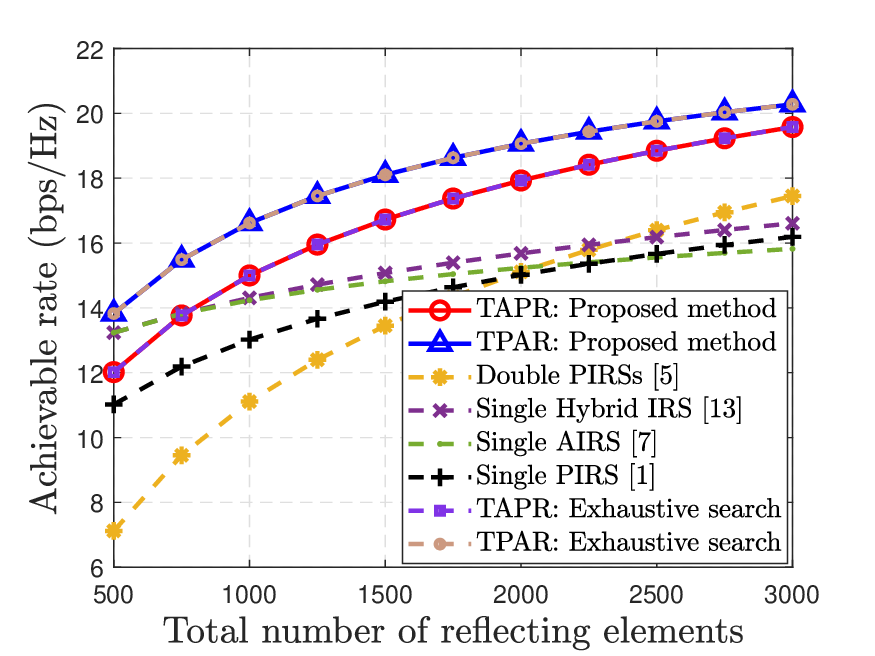}}
\subfloat[Achievable rate versus AIRS/PIR-\\S cost ratio.]{\label{ratio_vs_R}
		\includegraphics[scale=0.3]{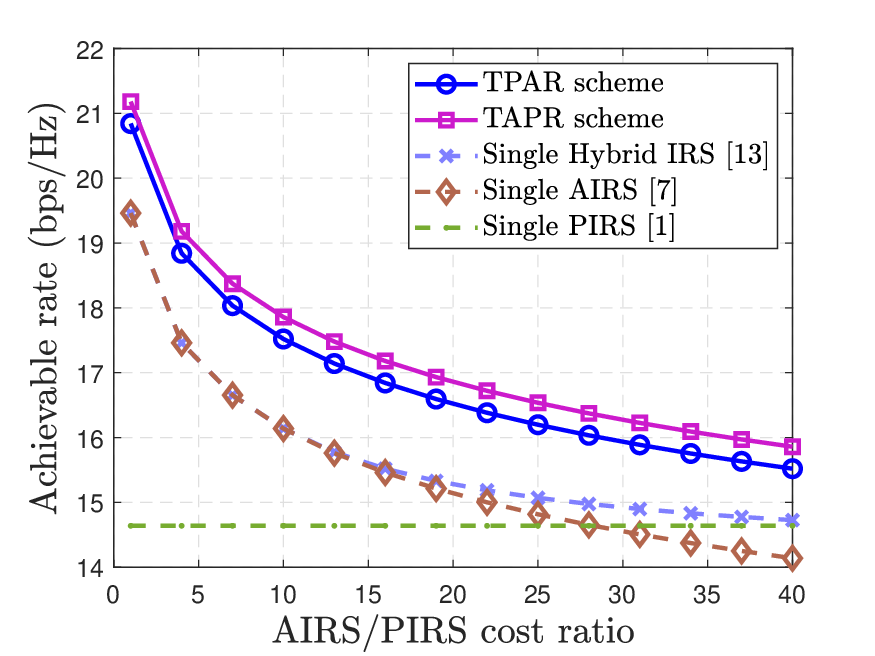}}
\caption{Rate comparison of joint AIRS-PIRS with benchmark systems.}
\label{fig3}
\end{figure}

In Fig. \ref{M_opt_sub}, we depict the number of A-IRS (i.e., the first IRS near Tx) elements versus the total element budget. It is observed that the proposed suboptimal solutions yield near-optimal designs, verifying the effectiveness of Lemma 1. In addition, the PIRS is equipped with more elements compared to AIRS in both schemes. In Fig. \ref{P_opt_sub}, we compare the achievable rate versus amplification power budget with $M\!=\!500$. It can be observed that with the growing amplification power budget, the rate performance of the TAPR system increases, while that of the TPAR system first rises very slightly and then tends to be constant. This is because the AIRS is closer to the Rx than TAPR and the amplification noise does not suffer from product-distance path loss, resulting in smaller amplification noise attenuation and greater SNR inhibition, and thus the growth of the achievable rate becomes smaller with the amplification power budget until the signal amplification power and amplification noise power reach dynamic equilibrium and the SNR tends to be constant. Note that if the distance between AIRS and Rx increases in the TPAR scheme, the growth magnitude of performance will be enhanced. Conversely, that of performance will be weakened. Moreover, when $P_v\! > \!\mathrm{15dBm}$, the TAPR scheme outperforms the TPAR scheme because it suffers less amplification noise power at the Rx and thus obtains better SNR. The result agrees well with Proposition 3. Finally, both TAPR and TPAR schemes achieve a better rate performance than double PIRSs case where two PIRSs satisfy optimal equal elements allocation thanks to the considerable amplification gain.


Fig. \ref{M_vs_R} shows the achievable rate of various systems versus the total number of reflecting elements with $P_v\! = \!\mathrm{10dBm}$. First, it is observed that TPAR scheme always outperforms single AIRS, single PIRS, and double PIRSs cases since the joint AIRS and PIRS architecture strikes a dynamic balance between the power amplification and beamforming gains. Furthermore, the TAPR scheme achieves higher performance than single hybrid IRS system when $M\! > \!750$. This is intuitively expected for the reason that with the adequately large array, the multi-IRS architecture provides an extra spatial degree for channel rank to control and reconfigure the wireless propagation environment and offers higher multiplicative beamforming gain. However, the hybrid IRS is easier to deploy and install as all active/passive reflecting elements are combined into a single IRS. In contrast, TAPR is lower than single hybrid IRS system when $M\! < \!750$ since under the small element budget, the amplification gain introduced by active elements is higher than passive-element beamforming gain. In addition, our considered architecture exhibits a greater growth rate than single PIRS and single AIRS thanks to the higher received SNR scaling order (i.e., $\mathcal{O}(M^3)$ versus $\mathcal{O}(M^2)$ and $\mathcal{O}(M)$). Lastly, we simulate the performance of the optimal integer solutions obtained by the exhaustive search method. It is observed that the rate performances of TAPR and TPAR with the active/passive elements optimized by the proposed method are close to those optimized by the exhaustive search algorithm, with a gap only in the third decimal place. In Fig. \ref{ratio_vs_R}, we investigate the achievable rate versus different ratios of active-over-passive deployment cost (i.e., $W_{\mathrm{act} }/ W_{\mathrm{pas} }$), where we set $W_{\mathrm{pas} }\!=\!1$, $M\!=\!1500$, and $P_v\!=\!17\mathrm{dBm}$. It can be obtained that our consider architecture has the best performance as compared to other benchmarks. Besides, with $W_{\mathrm{act} }$ increasing, the achievable rates of TAPR, TPAR, and hybrid IRS cases decrease initially, then gradually approach PIRS and tend to be unchanged when $W_{\mathrm{act} }$ is sufficiently large since the growth of $W_{\mathrm{act} }$ means that a smaller number of active elements can be deployed and a larger deployment budget can be allocated to passive elements.

\section{Conclusion}

In this paper, we studied elements allocation and beamforming design for maximizing the achievable rate in the AIRS-PIRS jointly aided wireless point-to-point communication system with two different deployment schemes (i.e., TAPR and TPAR). Under various practical constraints, we formulated the optimization problem, derived a series of element-related closed-form analytical expressions, and analyzed the performance of the two schemes. Moreover, we showed that the PIRS should be allocated more resources than the AIRS in both schemes and the received SNR increases asymptotically with the cube of the number of reflecting elements, when the distance between AIRS and PIRS is sufficiently large.
Numerical results were presented to validate our analysis and indicated that our considered architecture achieves higher rate performance than hybrid IRS and other existing benchmarks under the same deployment budget when $M$ is relatively large. In future work, it is an interesting direction to consider the energy efficiency with limited deployment budgets in this joint AIRS and PIRS system.


\end{document}